\documentstyle[epsf,epsfig,rotate]{mn}
 
\newcommand{\be}{\begin{equation}}
\newcommand{\ee}{\end{equation}}
\newcommand{\gs}{\;\raisebox{-.8ex}{$\buildrel{\textstyle>}\over\sim$}\;}

\newcommand{\apj}{{\it ApJ, }}

\newcommand{\mnr}{{\it MNRAS, }}

\newcommand{\ana}{{\it A\&A, }}

\newcommand{\MJup}{M$_J$}
       
\title
[Resonant planets in a disc]{Possible Commensurabilities Among 
Pairs of Extrasolar Planets} 
\author[R.P.Nelson \& J.C.B.Papaloizou]{Richard P. Nelson \&
John C.B. Papaloizou \\ 
Astronomy Unit, Queen Mary, University of London, Mile End Rd, London E1 4NS}
 
\date{Received/Accepted}

\begin{document}

\maketitle

\begin{abstract}
We investigate the possible
commensurabilities to be expected 
when two protoplanets in the Jovian
mass range, gravitationally interacting with each
other and an external
protoplanetary disc, are driven by disc induced 
orbital migration of the outer protoplanet
into a commensurability which is then
maintained in subsequent evolution. 
We find that for a variety
of protoplanet masses
and typical protoplanetary disc properties,  as well as the 
setting up of 2:1 commensurabilities of the type recently
observed in GJ876,
3:1 commensurabilities are often formed, in addition to
4:1, 5:1, and 5:2 commensurabilities which occur less frequently.
The higher order commensurabilites are favoured when
either one of the planets is massive, or the inner planet begins with
a significant orbital eccentricity.
Detection of such commensurabilities could yield
important information relating to the operation of protoplanet
disc interactions during and shortly post formation.

\end{abstract}

\begin{keywords} giant planet formation - extrasolar planets -
- orbital migration - resonance-protoplanetary discs - stars: individual: 
GJ876 
\end{keywords}

\section{Introduction} \label{intro}
The recent discovery of a pair of extrasolar giant planets orbiting 
in a 2:1 commensurability  
around GJ876 (Marcy et al 2001)
has raised interesting questions about the post-formation
orbital evolution of this system. Assuming the system did not form 
in its currently observed state, the existence of the commensurability
indicates that disc induced orbital migration is likely to have occurred 
so as to gradually reduce the planetary orbital
separation until resonance was established.

Simulations of two planets in the Jovian mass 
range interacting with a disc have been performed
by Kley (2000)  and Bryden et al (2000). 
The latter authors found a tendency
for the two planets to open up gaps in their local vicinity, and for material
between the two planets 
to be cleared, ending up interior to the inner
planet orbit or exterior to the outer planet orbit,
such that both planets orbit within an inner cavity.

Snellgrove, Papaloizou \& Nelson (2001) (hereafter paper I)
performed a 
simulation of a system consisting
of a primary star with two planets 
moving under their mutual gravitational attraction and forces produced by
tidal interaction with an 
externally orbiting gaseous disc. 
Angular momentum exchange caused the outermost planet to migrate inwards 
until a 2:1 commensurability with the inner planet was reached.
The subsequent dynamical interaction then resulted in the planets
migrating inwards together maintaining the commensurability.

The simulation results in paper I could be well matched with 
those of a simple $N$
body integration procedure used below, which incorporated
simple prescriptions for the migration and eccentricity damping of
the outer planet due to interaction with the disc.
Using this we investigate the range of commensurabilities
that might be expected in two planet systems as a consequence of protoplanet
disc interactions occurring in a standard protoplanetary disc model, where
we assume a fixed migration time and consider various eccentricity
damping rates.

Depending on the nature of the interaction between the disc and the 
outer planet, as well as the details of the resonance into which
the planets enter,
we distinguish between four types of orbital evolution
in the resonant migration phase which we denote as types A -- D. We
denote the orbital elements of the outer planet with a subscript `1'
and the inner planet with subscript `2'. Note that we find the evolution of
the system depends on the details of the resonance into which the planets
become locked. For higher order resonances such as 3:1 and 4:1, 
commensurabilities may be maintained in which differing resonant angles
librate  such that they do not 
cover the full $(0 , 2 \pi ) $ domain
(see for example Murray \& Dermott 1999 for an extensive discussion). 
For a $p:q$ commensurability,
we monitor the resonant angles defined by
\be \phi_{p,q,k}= p\lambda_1 -q\lambda_2 -p\varpi_1 +q\varpi_2
+k(\varpi_1 -  \varpi_2).\ee
Here $\lambda_1, \lambda_2, \varpi_1$  and 
$\varpi_2$ denote the mean longtitudes
and longtitudes of periapse for the planets $1$ and $2$
respectively.  The  positive integers $p$ and $q$
satisfy $p > q,$ and there are $p - q + 1$ possible values
of the positive  integer $k$ such that $ q \le k \le p.$
Although there are $p - q + 1$  corresponding angles,   no more than 
 two  can be linearly independent. 
This means that if libration occurs, either all 
librate or only one librates. Both situations occur in our 
integrations.
\\
{\bf Type A:} Eccentricities increase during migration until
they reach steady state equilibrium values, with possibly small oscillations 
superposed. After this migration continues in a self-similar manner
and all the resonant angles defined above are in libration. \\
{\bf Type B:} Eccentricities
increase as the migration proceeds, until 
$e_1 \ge 0.2$ at which point we estimate the outer planet
enters the outer disc. 
Our simple model breaks down at this point.
Large values of $e_1$ arise when the eccentricity damping rate is small.
As in type A resonant migration all the resonant angles considered 
go into libration. The difference is that $e_1$ exceeds $0.2$
 before a steady state  can be reached.
An experimental model of the evolution of a
two planet system corresponding to type B behaviour, and 
based on our knowledge of the migration of an eccentric planet
interacting with a protoplanetary disc is presented in section~\ref{e2_0_0.1}.
 \\
{\bf Type C:} For higher initial values of $e_2$,
we find a mode of evolution in which $e_1$ and $e_2$
rise continuously during the migration phase,
even for efficient damping of eccentricity. The issue
of the outer planet entering the disc again arises in type C migration. The 
primary difference between type C evolution and types A and B is that
only one of the resonant angles  defined above
is found to librate for the same apparent
commensurability, so that the  evolution differs. \\
{\bf Type D:} This mode of resonant migration corresponds to small
values of  $e_1 < 0.2 $ being  maintained for the outer planet, whilst the
eccentricity of the inner planet grows to attain values close to unity.
In this respect
it differs from the other types of migration. This mode of
evolution is related to that described by Beust \& Morbidelli (1996)
when discussing the generation of star grazing comets in the $\beta$ Pictoris
system.  All the resonant angles   librate in type D
migration but with large amplitude.

\section{Conservation of Energy and Angular Momentum}
We consider the consequences of a simple 
application of the conservation
of energy and angular momentum.
In the case when a near steady state for the orbital
eccentricities is attained, a relationship between the orbital eccentricities 
and the circularization and migration rates induced
by the disc is obtained for the case
of type A migration.

We consider  two planets with masses $m_1, m_2,$ osculating
semi-major axes $a_1, a_2,$  and eccentricities $e_1 ,e_2.$
These orbit a central mass $M_*.$
The total angular momentum is
\be J= J_1+J_2 = m_1\sqrt{GM_* a_1(1-e_1^2)} +  m_2\sqrt{GM_* a_2(1-e_2^2)}\ee
and the energy $E$ is
\be E = -{GM_* m_1\over 2 a_1} -{GM_* m_2\over 2 a_2} \ee
We assume  resonant self-similar  migration in which $a_2/a_1,$ $e_1,$ and $e_2$
are constant. Then conservation of angular momentum gives
\be {dJ\over dt} =J_1{1\over 2  a_1}
{d a_1\over dt}\left( 1 +
{m_2\sqrt{a_2(1-e_2^2)}\over m_1\sqrt{a_1(1-e_1^2)}}\right)
=-T \label{CJ} \ee
and conservation of energy gives

\be {dE\over dt} ={GM_* m_1\over 2 a_1^2} {d a_1\over dt}
\left(1+{m_2 a_1\over m_1 a_2}\right)
=-{n_1T\over \sqrt{1-e_1^2}} - D , \label{CE}\ee
where $n_1= \sqrt{GM_*/a_1^3}$ and 
we suppose there is 
a tidal torque $-T$ produced by the disc
which acts on $m_1.$
 In addition we suppose there to be
 an associated
 tidally induced orbital energy  loss rate
 which is  written as  $n_1 T/\sqrt{1-e_1^2} +D,$ 
 with 
$D \equiv (GM_* m_1 e_1^2) /(a_1(1-e_1^2) t_c).$
Here $t_c$ is the circularization time of $m_1$
that would apply if the tidal torque and energy
loss rate acted on the orbit of $m_1$ with $m_2$ being absent.
In that case we would have $de_1/dt = -e_1/t_c.$
A migration time $t_{mig}$ can be defined through
$T= m_1\sqrt{GM_* a_1(1-e_1^2)}/(3 t_{mig}).$
This is the time for $n_1$ to increase by a factor of
$e$ if  $m_2$  was absent and the
eccentricity $e_1$ was fixed. Note that $t_c$ and $t_{mig}$
are determined by the disc planet tidal interaction
and may depend on $e_1.$

\noindent By eliminating ${d a_1\over dt}$ from (\ref{CE}) and (\ref{CJ})
we can obtain a relationship between $e_1, e_2, t_c,$ and $t_{mig}$ in the form
\be e_1^2 = 
{t_c \left(1-e_1^2
-{(1-e_2^2)^{1/2}(1-e_1^2)^{1/2}\over a^{-3/2}_2 a_1^{3/2}}\right)
\over 3 t_{mig}\left( {m_1a_2\over m_2a_1}  +
{a_2^{3/2}(1-e_2^2)^{1/2}\over a_1^{3/2}(1-e_1^2)^{1/2}}\right)}
\label{ejcons} \ee

We comment that this is general in that it depends only on the conservation laws
and applies for any magnitude of eccentricity.  However, it's derivation did 
assume self-similar migration with  equilibrium eccentricities 
and so it does not apply in non equilibrium situations 
for which the eccentricities grow continuously. Note too that
for a 2:1 commensurability for which $(a_1/a_2)^{3/2}=2,$ (\ref{ejcons}) reduces
in the case $e_i^2 \ll 1$ 
to the expression given in paper I which was obtained by  analysis of the perturbation
equations directly: 

\be e_1^2 = {t_c m_2a_1\over 3 t_{mig}(2 m_1 a_2 + m_2a_1)} \label{ec1} \ee
The above determines the eccentricity of the outer planet
$e_1$ as a function of $t_c$ and $t_{mig}.$

\section{Numerical Calculations}
\subsection{Model Assumptions and Physical Setup}
The basic assumptions of our model
are that the two planets orbit
within the inner cavity of a tidally truncated disc that lies
exterior to the outer planet.
Tidal interaction with this disc causes inwards migration of the outer planet
on a time scale of $t_{mig}$,
and also leads to eccentricity damping of the outer planet on a time scale of
$t_c$. We assume that the inner edge of the disc is such that an eccentricity
of $e_1 \; \gs \; 0.2$ will enable the outer planet to enter the disc, in basic agreement with the results of hydrodynamic simulations 
(e.g. Nelson et al 2000).

We have performed
three-body orbit integrations using a fifth-order Runge-Kutta scheme,
and the Burlisch-Stoer method as an independent check
(e.g. Press et al. 1993). 
A torque was applied
to the outermost planet such that it migrated inwards on a time scale
of $t_{mig}= 10^4 $ local orbital periods
and a damping force proportional to the radial velocity
was applied in
the radial direction.   This has a fixed
constant of proportionality such that for small $e_1$
and in the absence of disc torques, 
 the eccentricity  damps on a  time scale of
$t_c=600{\cal N} $ local orbital periods with values of
${\cal N} = 1,10,100$ having been considered.
We comment that 
we have adopted the simplification of neglecting
any possible dependence of 
the damping force and  
 excepting  the calculation
presented in section \ref{e2_0_0.1},
 $t_{mig},$ on $e_1.$ 

This procedure was shown to be capable of matching the results of
a detailed simulation in paper I, and
the value of $t_{mig}$ adopted is consistent with that found
from hydrodynamic simulations of protoplanetary discs
interacting with protoplanets in the Jovian mass range (e.g. Nelson et al 2000).
Consideration of the orbital parameters of GJ876 (see paper I)
suggests $ {\cal N} \sim 1.$ However, there is some uncertainty
in  the functional dependence of the  circularization rate of an orbiting 
protoplanet nonlinearly interacting
with a tidally truncated protostellar disc as it would
depend on the distance to and form of the disc edge 
(Goldreich \&  Tremaine 1980),
and be sensitive to the presence of coorbital material
which can lead to eccentricity damping through coorbital Lindblad torques
(Artymowicz 1993).  

For length scale we adopt
a fiducial radius, $R$, which for simplicity is taken to be 
1 AU. However, scaling invariance allows this to be scaled to some 
other value if required.
Runs were typically started with the outer planet
at radius $15.6R$ while the inner planet was started at radius $5R.$
A larger initial separation would be inconsistent with our model assumption
of there being two planets orbiting within a tidally cleared cavity. Simulations
by Bryden et al. (2000) suggest that the time scale to clear such a cavity 
becomes comparable to the migration time if the ratio of planetary semimajor
axies becomes much larger than 3.
The outer planet was assumed, because of interaction with the disc,
to be in a circular orbit while
starting  eccentricities  $e_2 =0.0, 0.05, 0.1, 0.2, 0.3$ were
adopted for the inner planet. 
For protoplanet masses we have considered
all permutations of $0.4,1$ and $4$
Jupiter masses
assuming the central star mass is 1 M$_{\odot}$.

\subsection{Results}
 
 \begin{table}
 \begin{center}
 \begin{tabular}{|l|l|l|l|l|l|l|l|} \hline \hline
 $m_1$ & $m_2$ & $\cal{N} $ & &  Resonance & & &   \\
 \hline
      &       &               &$e_2=0$& 0.05&0.1&0.2 & 0.3  \\
      \hline
      4 & 4&  1 &3:1A &3:1A & 3:1A  &4:1A &3:1A \\
      4 & 4&  10 &3:1B &4:1B & 4:1B  &3:1B &4:1C\\
      4 & 4&  100&3:1B &3:1B & 3:1B  &3:1B &5:1C\\
      4 & 1&  1 & 3:1A &3:1A & 4:1A  &3:1A &2:1A\\
      4 & 1&  10 & 3:1B &3:1B & 3:1B  &2:1A &4:1D\\
      4 & 1&  100 &3:1B & 3:1B & 3:1B  &3:1C &2:1B\\
      4 & 0.4& 1 & 3:1A &3:1A& 3:1A  &5:2A &4:1A\\
      4 & 0.4& 10 &3:1A & 3:1A& 3:1A  &3:1A &2:1A\\
      4 & 0.4& 100 & 3:1B& 3:1B & 3:1B &2:1B &2:1B\\
      \hline
      1 & 4&  1 & 2:1A& 3:1A& 3:1A &3:1A &3:1C\\
      1 & 4&  10 & 3:1B& 3:1B& 3:1B &5:1C &5:1C\\
      1 & 4&  100  & 3:1B& 3:1B& 4:1B  &5:1C &3:1B\\
      1 & 1&  1 & 2:1A& 2:1A & 3:1A  &4:1C &2:1A\\
      1 & 1&  10 & 2:1B& 2:1B & 2:1B  &4:1B &5:1B\\
      1 & 1&  100  & 2:1B& 3:1B & 4:1B  &3:1B &5:1B\\
      1 & 0.4& 1 &2:1A & 3:1A & 3:1A  &2:1A &2:1A\\
      1 & 0.4& 10 &2:1B & 2:1B & 3:1B  &3:1B &3:1B\\
      1 & 0.4& 100 & 2:1B& 3:1B & 3:1B  &2:1B &3:1B\\
      \hline
      0.4 & 4& 1 &2:1A & 3:1A & 3:1A &3:1C &3:1C\\
      0.4 & 4& 10 & 2:1B& 3:1B & 3:1B &5:1C &5:1C\\
      0.4 & 4& 100 & 3:1B& 3:1B & 4:1B &5:1C &5:2B\\
      0.4 & 1& 1 &2:1A & 2:1A & 3:1A &2:1A &4:1C\\
      0.4 & 1& 10 & 2:1B& 3:1B & 2:1B  &2:1B &5:2C\\
      0.4 & 1& 100 &2:1B & 5:2B & 3:1B &5:2C &2:1B\\
      0.4 & 0.4& 1& 2:1A& 2:1A & 3:1A & 2:1A&2:1A\\
      0.4 & 0.4& 10 &2:1B & 2:1B & 2:1B &3:1B &3:2B\\
      0.4 & 0.4& 100 &2:1B & 3:1B & 2:1B &2:1B &2:1B\\
       
       \hline
       \end{tabular}
       \end{center}
       \caption{This table indicates the outcome
       and commensurability attained when a pair of masses $m_1, m_2$
       measured in Jupiter masses
       becomes resonantly coupled due to disc driven migration
       of the outer planet. The initial eccentricity $e_2$ and $\cal{N}$
       are indicated, $e_1$ is always initated as zero. The letters
       A, B, C, and D in columns 4 -- 8 indicate whether types A, B, C, or D
       evolution occurred.}
       \label{tab1}
       \end{table}

\begin{figure}
\epsfig{file=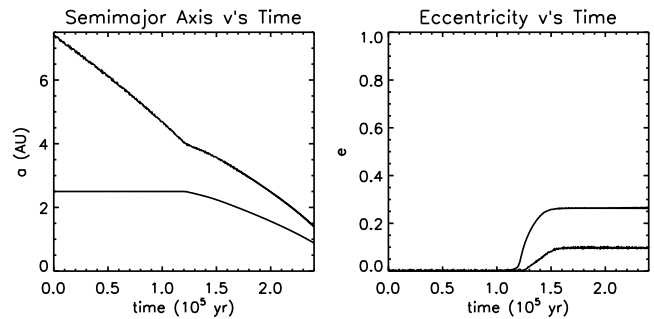,width=\columnwidth}
\caption{This figure shows the evolution of the planet
semimajor axes and
eccentricities for a pair
of protoplanets with $m_1=1$, $m_2=1$, initial $e_2=0$, and
${\cal N}=1.$
This case underwent self-similar (type A) migration in 2:1 
resonance after having attained
equilibrium eccentricities.
The outer planet is denoted by the upper line in the first panel
and the lower line in the second panel. 
The unit of time is $10^5 (R/{\rm 1AU})^{3/2}$ yr.}
\label{fig1}
\end{figure}  

\begin{figure}
\epsfig{file=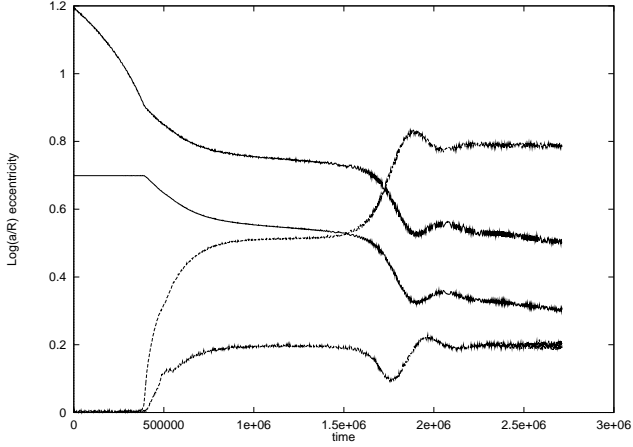, angle =270, width=\columnwidth}
\caption
{This figure shows the evolution of the  protoplanet
semimajor axes and
eccentricities for a pair
of protoplanets, each of 1 Jupiter mass  with
${\cal N}=100$ but with the migration rate of the outer planet
modified such that it reverses for $e_1>0.2.$
At small times, the upper two curves show $\log_{10} (a/R)$ while
the lower two curves show $e_1$, $e_2$. The upper curve of the two denoting
$\log_{10} (a/R)$ represents the outer planet, whereas at late times
the lower curve of the two representing $e_1$, $e_2$ represents the outer 
planet.
The unit of time is $(R/{\rm 1AU})^{3/2}$ yr.}
\label{fig2}
\end{figure} 

\begin{figure}
\epsfig{file=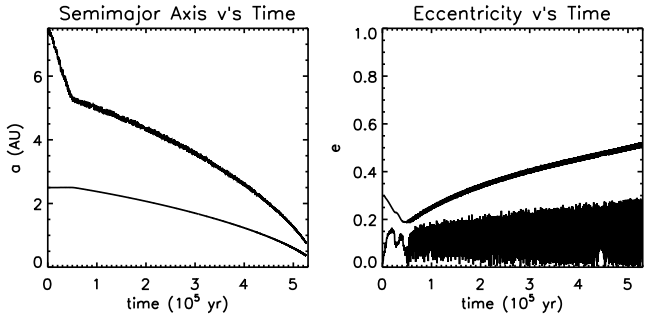, width=\columnwidth}
\caption{This figure shows the evolution of the protoplanet
semimajor axes and eccentricities for a pair of protoplanets,
with $m_1=1$, $m_2=4$, initial $e_2=0.3$, and ${\cal N}=1$. This case 
provides an example of type C migration in 3:1 resonance. Here the 
eccentricities of both planets grew  continuously as the
system evolved, even with significant damping of $e_1$ by the disc.
Note that our simple model breaks down for $e_1 > 0.2$.
The outer planet is denoted by the upper line in the first panel and
the lower line in the second panel. 
The unit of time is $10^5 (R/{\rm 1AU})^{3/2}$ yr.}

\label{fig3}
\end{figure}

\begin{figure}
\epsfig{file=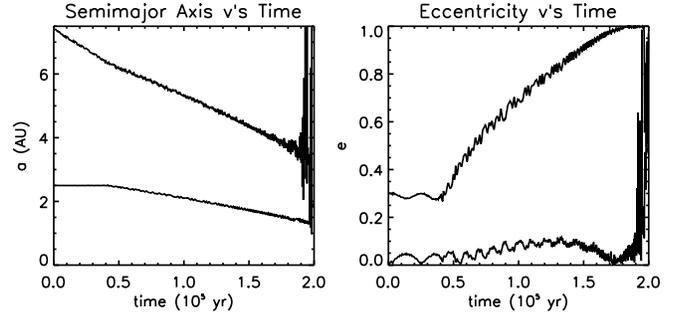, width=\columnwidth}
\caption
{This figure shows the evolution of the protoplanet
semimajor axes and
eccentricities for a pair
of protoplanets, with $m_1=4$, $m_2=1$, initial $e_2=0.3$, and
${\cal N}=10$.
This run provides an example of type D migration in 4:1 resonance.
Here the eccentricity of the inner planet grows without limit during the
evolution, eventually reaching $e_2 \simeq 1$, whereas the eccentricity
of the outer planet remains relatively small. As $e_2$ approaches 1, a
dynamical instability occurs, causing the planets to be scattered.
The simplicity of our disc model does not allow this phase of evolution to 
be modeled accurately. The outer planet is denoted by the upper line in the 
first panel, and by the lower line in the second panel.
The unit of time is $10^5 (R/{\rm 1AU})^{3/2}$ yr.}
\label{fig4}
\end{figure}

The results of our survey are shown in table~1.
In general for the shortest circularization rates with ${\cal N}=1$
and initial values of $e_2 \le 0.1$, 
self-similar (type A) migration occurred. 
For larger values of ${\cal N}$, type B migration was usually the preferred
outcome.

For larger initial values of $e_2$ and ${\cal N}=1$, either type A or type C
migration occurred. For larger values of ${\cal N}$, types A, B, C, or D
were all found to occur, with a strong preference for types B and C.
Only one example of type D migration arose in our simulations.

\subsection{Initial $e_2=0$}
For an inner planet on an initially circular orbit,
a 2:1 commensurability normally resulted,
except when one of the masses was $4$ Jupiter masses in which case
a 3:1 commensurability could occur. 
Plots of the evolution of the semimajor axes and eccentricities
of a Jupiter mass pair with ${\cal N}=1$  which undergo self-similar
migration are illustrated in figure \ref{fig1}.
In this case steady equilibrium eccentricities
are obtained that are
consistent with equations
(\ref{ejcons}) and (\ref{ec1}).
We found that for the adopted migration rate, the transition 
between a 2:1 and a 3:1 commensurability occured
for $m_1$ in the range  2--3 Jupiter masses when $m_2$ was
a Jupiter mass, with the transition mass being smaller
for larger migration rates.

\subsection{ Initial $0 < e_2 \le 0.1$} \label{e2_0_0.1}
When the inner planet was started with an initial
eccentricity such that $0 < e_2 \le 0.1$,
trapping in a 3:1 commensurability was more common and could
occur even for the lowest mass pairs. Also a few cases
of trapping in a 4:1 commensurability and one 5:2 commensurability
were found. It was observed that the mode of migration
in these cases usually corresponded to type A for ${\cal N}=1$, and type B
for ${\cal N}=10$ or 100. 

In order to investigate the potential outcome of a type B 
migration we considered a Jupiter mass pair with ${\cal N}=100$
in which case the eccentricity of the outer planet grows to exceed
$0.2.$ At this stage the assumed migration rate becomes inapplicable.
In fact migration 
may reverse on account of penetration
of the disc by the outer planet. In this context we note that a simulation
of massive protoplanets by Papaloizou, Nelson \& Masset (2001)
indicated such a potential reversal for $e_1 > 0.2$
and linear torque calculations by Paploizou \& Larwood (2000)
suggested migration  reversal for an embedded protoplanet with eccentricity
exceeding a few times the disc aspect ratio. As an experiment
we modified the disc induced migration rate of $m_1$ by a factor
$(1- (e_1/0.2)^2).$
The resulting evolution is shown in figure~\ref{fig2}
This adjustment can stall the migration rate
as $e_1$ approaches $0.2$ and lengthen the lifetime
of the system before a potential
orbital instability occurs due to the development of large
eccentricities in both planets. In fact in this case the eccentricity
of the inner planet reached $\sim 0.8$ after $\sim 2\times 10^6
(R/{\rm 1AU})^{3/2}$~y.
The system was found to be subsequently stable for at least a similar
time after a rapid disc removal.

Thus we emphasise that the projected lifetimes of the resonantly migrating
systems discussed here are dependent on the nature of the disc
interaction, and at present are very uncertain. However, they could
approach the protoplanetary disc lifetime if the mode of evolution discussed
in the preceding paragraph was to occur.

\subsection{Initial $0.2 \le e_2 \le 0.3$}
For initial values of $e_2$ in the range $0.2 \le e_2 \le 0.3$, it was observed
that type C evolution became an important consideration, with only a single
case of type D evolution being observed.
For type C evolution, as in type B, the eccentricities of both planets
were observed
to grow without reaching equilibrium, until $e_1 > 0.2$ at which point our
simple model breaks down. The long term evolution in this situation
may resemble that described in figure~\ref{fig2}. We illustrate the
behaviour of type C evolution in figure~\ref{fig3}.

The single case of type D evolution obtained is illustrated in 
figure~\ref{fig4}, which should be compared with figure~\ref{fig3}
since the only difference between these runs is the initial value of $e_2$
($e_2=0.2$ in figure~\ref{fig3} and $e_2=0.3$ in figure~\ref{fig4}).
In this mode of evolution, the outer planet maintains 
a modest eccentricity while the eccentricity of the inner planet
grows without limit. Although this mode of evolution appears to be
rather rare based on the results of our calculations, it is interesting to
note that it provides a method of generating very high eccentricities,
such as that observed in the system HD80606 where the planetary
eccentricity is $e \; \gs \; 0.9$ (Naef et al. 2001).

Even for initially high values of $e_2$, we still find a tendency for
lower mass pairs of planets to enter lower order resonances such as 2:1, 
3:1 and
even 3:2, whilst the higher mass pairs tend to occupy the higher order
commensurabilities, even resulting in capture into 5:1 on a significant number
of occassions.

\subsection{The Case of GJ876}
We have run a number of calculations to examine the putative early evolution
of the observed system GJ876, which exists in a 2:1 commensurability.
The minimum masses of the two planets, in units where the central star
s of solar mass, are $m_1=6$ \MJup and $m_2=1.8$ \MJup. These parameters
suggest that trapping in higher order resonance may have been
expected rather than in 2:1. For initial conditions in which 
$m_1=6$, $m_2=1.8$, initial $e_2=0$, capture into 3:1 resonance is favoured.
For initial $e_2=0.1$, 4:1 resonance results.
However, in a scenario in which the planetary masses were smaller
during the initial resonant capture, capture into 2:1 can occur.
Simulations were performed with $m_1=1$, $m_2=1.8$, initial $e_2=0$ and 0.1,
which all favour capture into 2:1. Assuming a mass for the outer
planet of $m_1=3$ produces capture in 3:1.

Two possible conclusions to draw from this are that: \\
(1). resonant capture occurred in this system when the planets were
of smaller mass, and that the mass of the outer planet increased due to
accretion of gas from the disc.\\
(2). The planets formed sufficiently close together initially that
 capture into resonances of higher order than 2:1 was not possible.

\section{Summary and Discussion} \label{blah}
In this paper we have
considered two protoplanets gravitationally interacting with each
other and a protoplanetary disc. The two  planets orbit
interior to a tidally maintained disc cavity
while the disc interaction induces
inward migration. 
We find that as well as 2:1 commensurabilities being formed,
3:1 commensurabilities are obtained in addition to
4:1, 5:1, and 5:2 commensurabilities which occur less frequently and usually
for larger initial eccentricities of the inner planet.
Possession of a small eccentricity $\sim 0.1$  before resonant locking takes
place would favour 3:1 commensurabilities if migration
over large separations has occurred. 
Initially near circular orbits would favour 2:1 commensurabilities
except when one of the planets is significantly more massive than one Jupiter
mass.

Up to now  seven multiple planet systems have been detected
with two showing a 2:1 commensurability (GJ876, HD82943 -- see the website: \\
obswww.unige.ch/$\sim$udry/planet/new\_planet.html).
It is too early at present
to establish 
the frequency of occurence of commensurabilities in multiple systems
but they  may be fairly common.
If orbital migration has occured over a large scale
in extrasolar planetary systems the future detection
of   additional commensurabilities of the type discussed here 
is to be expected.
We also comment that in some cases a high eccentricity for the
inner planet orbit may be produced through resonant migration.
If, depending on the details
of disc dispersal and interaction,
a scattering followed by ejection of the outer planet
occurs, a single planet remaining on an eccentric orbit may result.
This will be a topic for future investigation.


\centerline{\bf {Acknowledgements}}

\noindent This work was supported by
PPARC grant number PPARC GR/L 39094.

\end{document}